\documentclass{elsart}

\usepackage{amssymb,graphicx}

\begin{document}

\begin{frontmatter}



\title{High-$p_T$ Strange Particle Spectra and Correlations in STAR}

 \author{Jana Bielcikova for the STAR Collaboration}
 \address{Physics Department, Yale University, New Haven, CT 06520, USA}

\begin{abstract}
 We present results on strange particle production in p+p, d+Au and Au+Au collisions
at $\sqrt{s_{NN}}$~=~200~GeV. We discuss the nuclear modification factors ($R_{CP}$) and
ratios of particle yields in order to investigate the range of anomalous baryon production.
The medium modification of the fragmentation process is further explored by studying
azimuthal correlations at intermediate-$p_T$ for $K^0_S$,
$\Lambda$ and $\bar{\Lambda}$ trigger particles. The results are compared
to fragmentation and recombination models.

\end{abstract}


\end{frontmatter}

The observed suppression of inclusive $p_T$ spectra of identified
particles in central Au+Au collisions with respect to p+p ($R_{AA}$) and
peripheral Au+Au collisions ($R_{CP}$)~\cite{Adler:2003kg,Adams:2003am}, together
with enhanced {\it baryon/meson} ratios~\cite{Adams:2006wk}, show that
in the intermediate-$p_T$ range of 2-6~GeV/$c$, baryons and mesons behave
differently than in p+p collisions. This indicates that jet fragmentation
is not a dominant source of particle production and parton recombination and
coalescence models have been suggested as alternative
mechanisms~\cite{Fries:2003kq,Greco:2003mm,Greco:2003xt,Hwa:2002tu}.
In addition, studies of two-particle correlations of charged particles in Au+Au
revealed the presence of additional long-range pseudo-rapidity correlations
on the near-side (commonly referred to as the {\it ridge})~\cite{PutschkeHP}. These are absent in p+p and d+Au collisions.

The wealth of data collected by the STAR experiment in p+p, d+Au and
Au+Au collisions at $\sqrt{s_{NN}}$~=200~GeV allows 
a detailed study of strangeness production up to high $p_T$.
In this paper, we discuss the behaviour of strange baryon/meson ratios and identified particle $R_{CP}$
in order to investigate the range of the baryon  anomaly. We further explore the medium
modification of the fragmentation process by studying two particle azimuthal correlations
at intermediate-$p_T$ using strange trigger particles ($K^0_S$, $\Lambda$, $\bar{\Lambda}$)
associated with unidentified charged particles.

\begin{figure}[t!]
\begin{tabular}{lr}
\includegraphics[height=5.5cm]{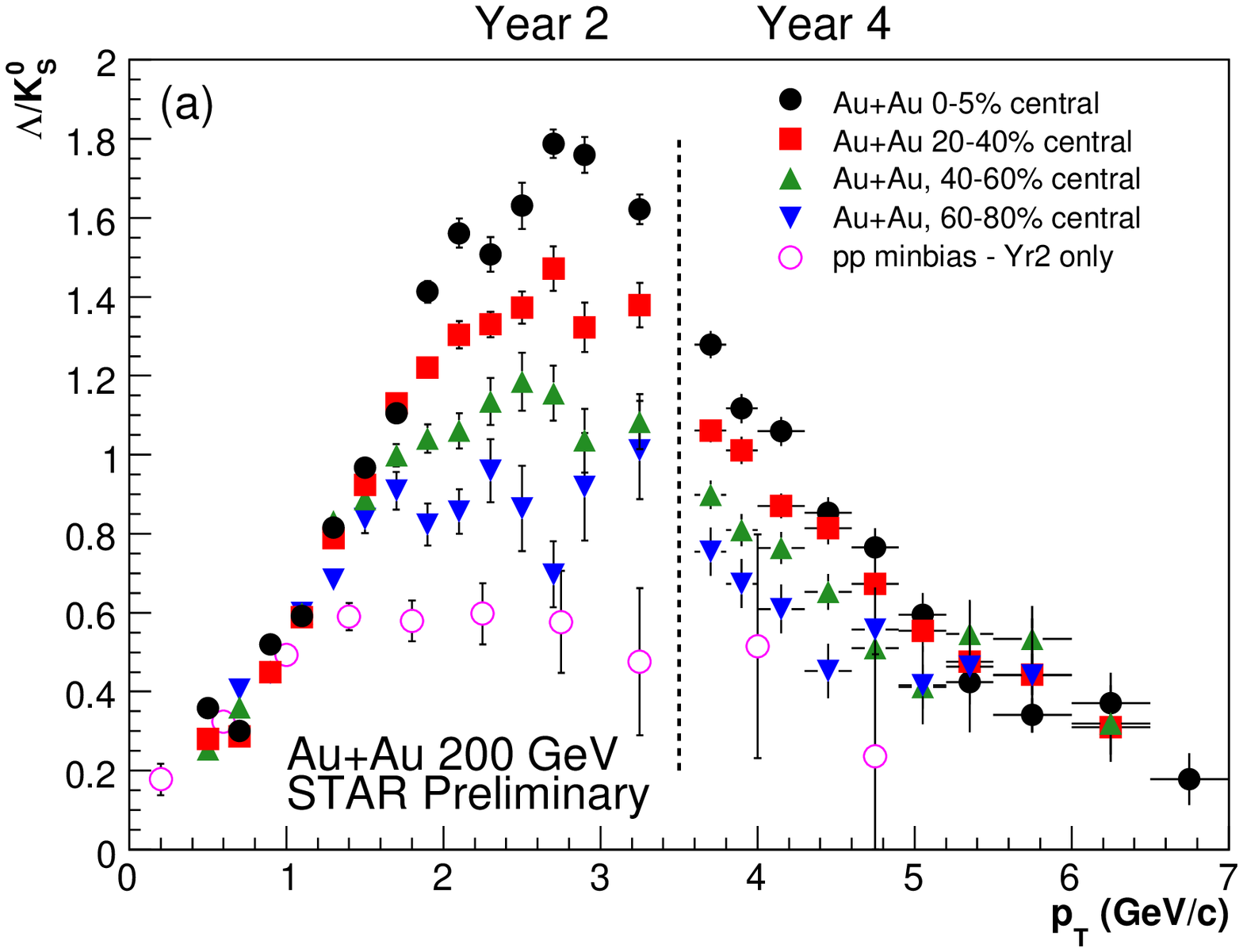}
&
\hspace{-0.5cm}
\includegraphics[height=5.5cm]{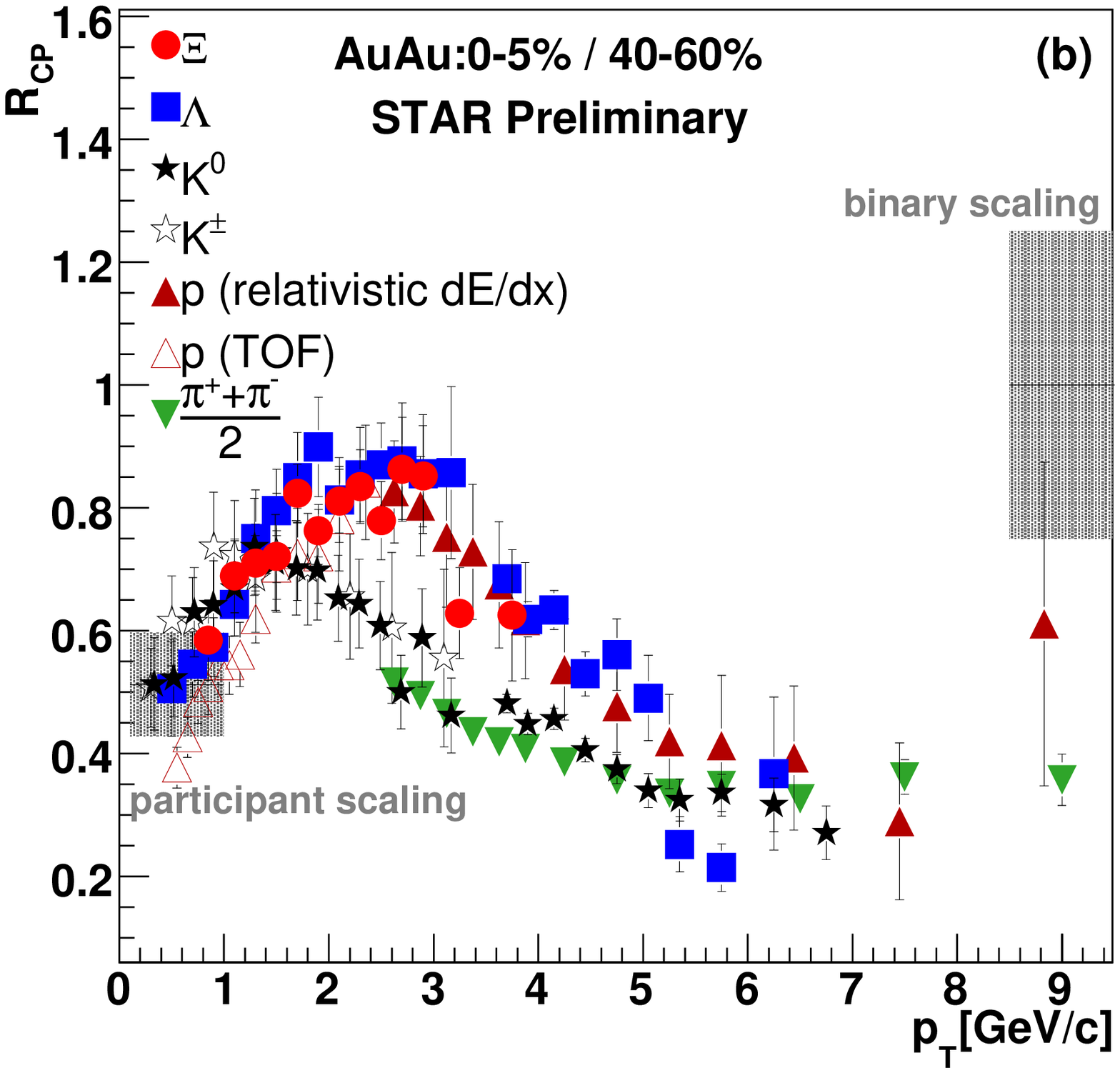}
\end{tabular}
\caption{(a) $\Lambda/K^0_S$ ratio as a function of $p_T$ in p+p and for different centralitites in
Au+Au collisions. (b) $R_{CP}$ of strange and non-strange particles in Au+Au collisions.}
\label{ratiorcp}
\end{figure}

Figure~\ref{ratiorcp}(a) shows the $\Lambda/K^0_S$ ratio for various centralities in Au+Au collisions.
The high statistics Year 4 data significantly increase the reach in $p_T$ and we observe
that by $p_T$~=~6~GeV/$c$, the ratio becomes similar for all centralities and is consistent with that in p+p.
Similarly, the $R_{CP}$ of identified particles (Fig.~\ref{ratiorcp}(b)), demonstrates that in
the intermediate-$p_T$ region mesons have smaller $R_{CP}$ than baryons,
but above $p_T$~=~6~GeV/$c$, they both show the same suppression.
Recombination/coalescence models predict a turn-over in the $\Lambda/K^0_S$ ratio,
however its exact position and shape of the distribution are not reproduced well~\cite{Fries:2003kq,Greco:2003mm,Greco:2003xt,Hwa:2002tu}.
The model combining soft particle production at low $p_T$ with a leading order pQCD at high $p_T$ incorporating gluonic baryon junctions and GLV formalism~\cite{Gyulassy:2000fs} predicts the turn-over in the ratio as well, but also fails to describe its position and shape in detail~\cite{Vitev:2001zn}.




In the following paragraphs we discuss properties of the near-side azimuthal
correlations using identified strange trigger particles ($\Lambda$, $\bar{\Lambda}$, and K$^0_S$)
associated with charged particles. The azimuthal distributions, normalized to the number of trigger particles, are corrected
for the reconstruction efficiency of associated particles and elliptic flow.
The near-side yield of associated particles is calculated as the area under
the Gaussian peak obtained from the fit.
We study separately the jet and ridge contributions to the near-side yield
by analyzing the correlations in two $\Delta\eta$ windows:
$\Delta\eta<$~0.5 containing both jet and ridge contributions, and $\Delta\eta>$~0.5
containing the ridge-like correlations.

Comparing d+Au and Au+Au collisions, we observe an increase of the near-side yields
by a factor of 3-4 going from d+Au to central Au+Au collisions.
While the jet yield is independent of centrality,
the ridge yield is responsible for the strong increase
of the total near-side yield (not shown here, see~\cite{BielcikovaHQ}).
No significant baryon/meson or particle/anti-particle differences are observed.
\begin{figure}[t!]
\begin{center}
\begin{tabular}{lr}
\hspace{-1.2cm}
\includegraphics[height=6.0cm]{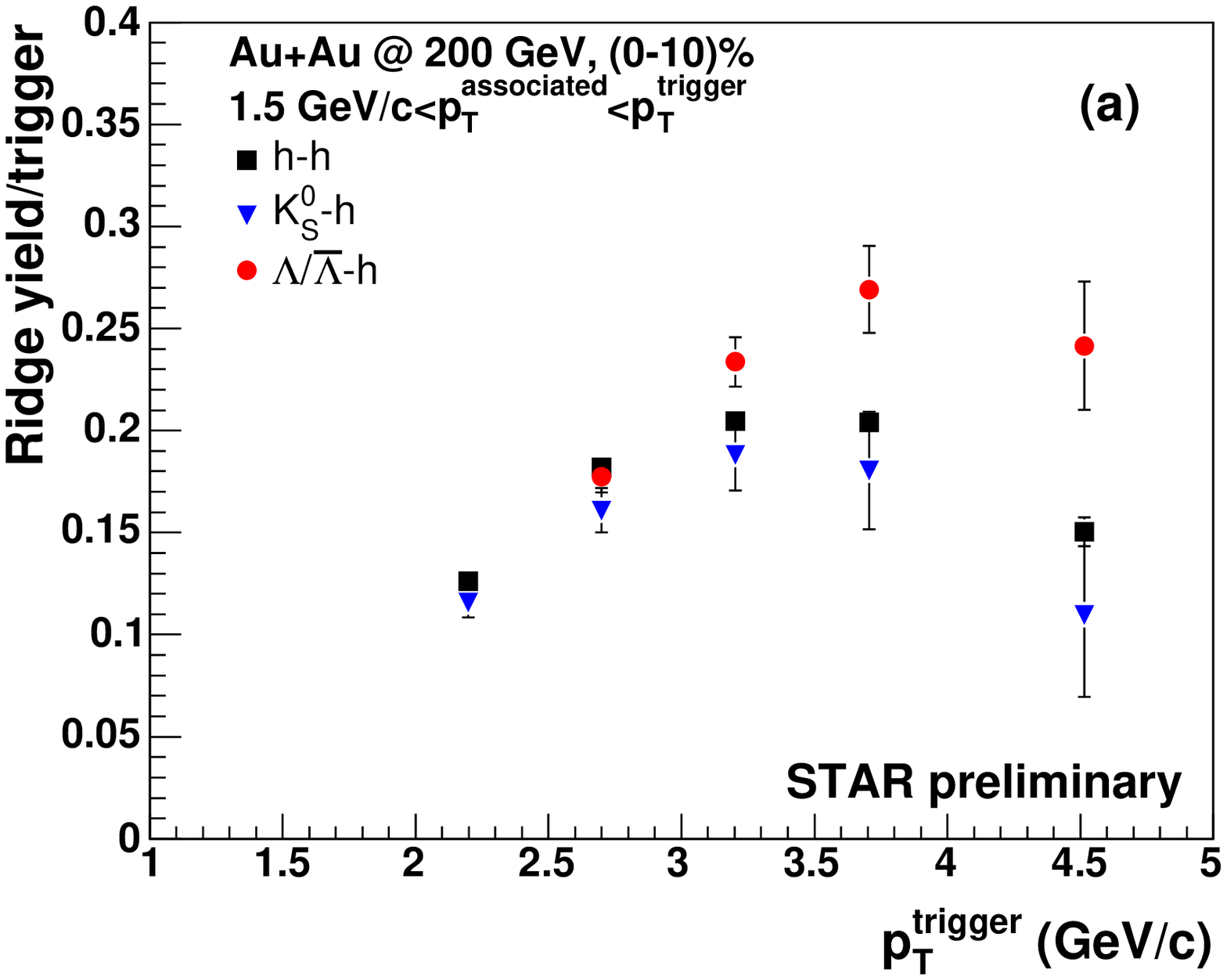}
&
\hspace{-0.8cm}
\includegraphics[height=6.0cm]{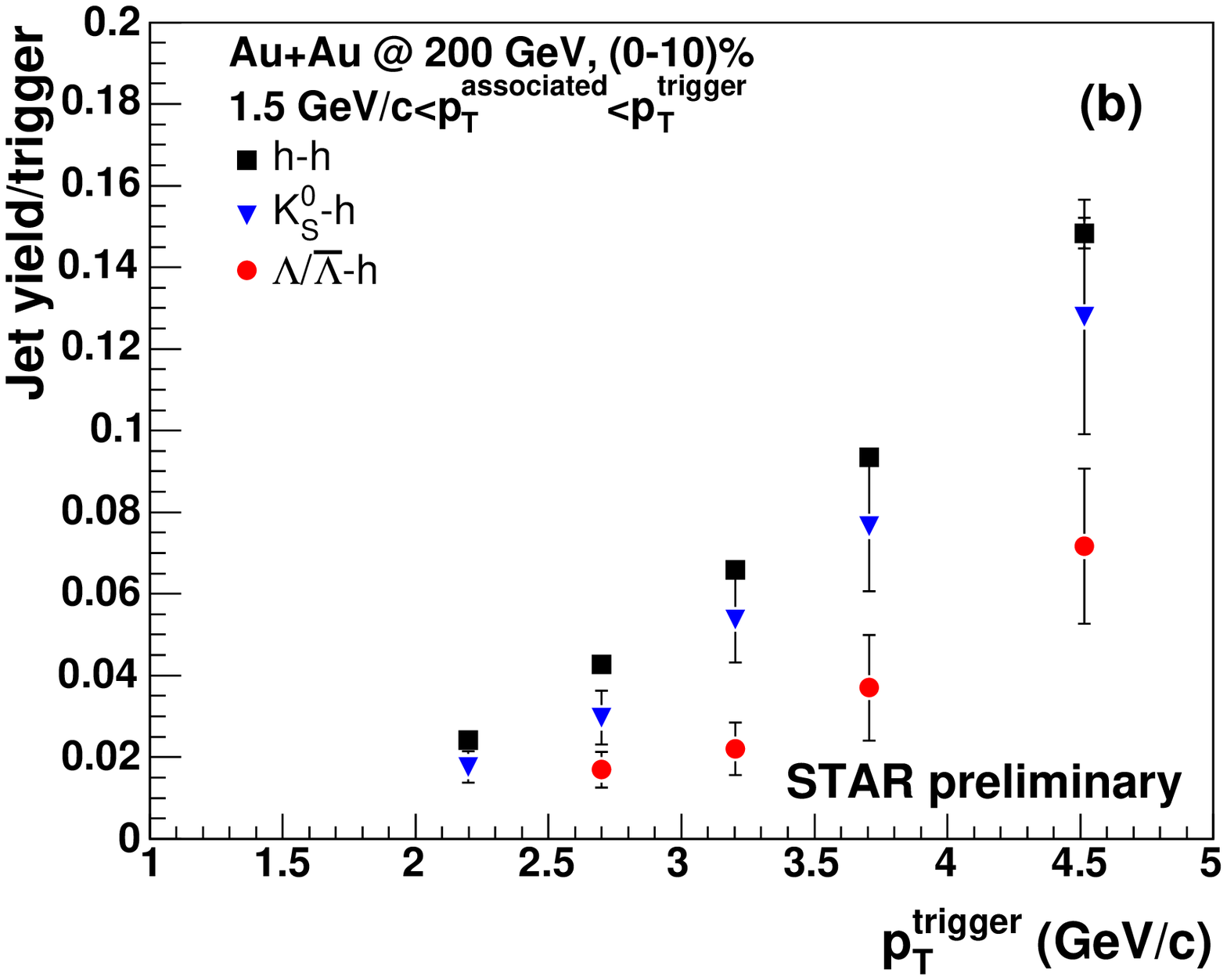}
\end{tabular}
\end{center}
\caption{Dependence of the ridge yield (a) and jet yield (b) on $p_T^{trigger}$ for various trigger species
in central (0-10\%) Au+Au collisions.}
\label{rjpttrig}
\end{figure}
\begin{figure}[b!]
\begin{tabular}{lr}
\includegraphics[height=6.5cm]{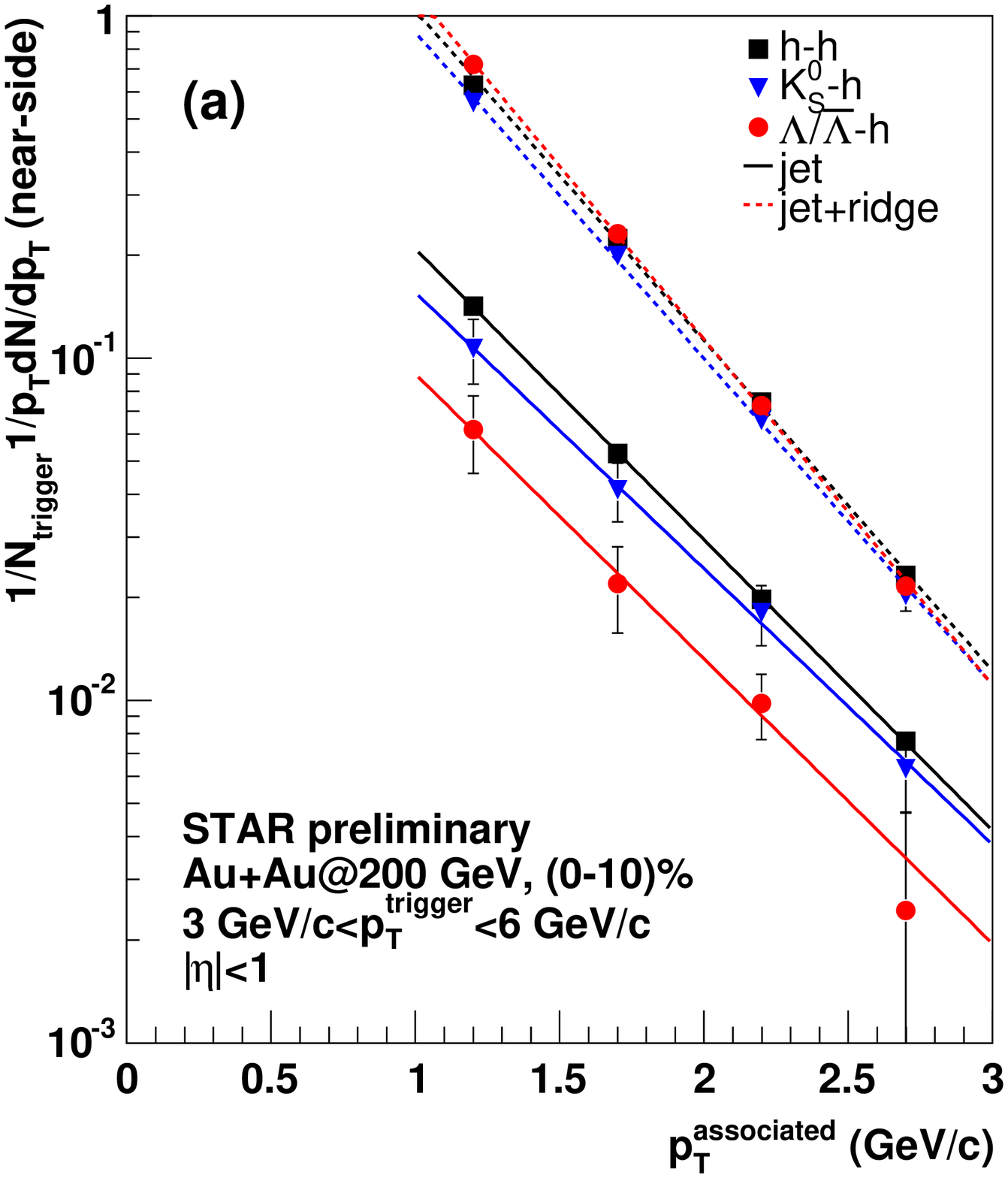}
&
\includegraphics[height=6.5cm]{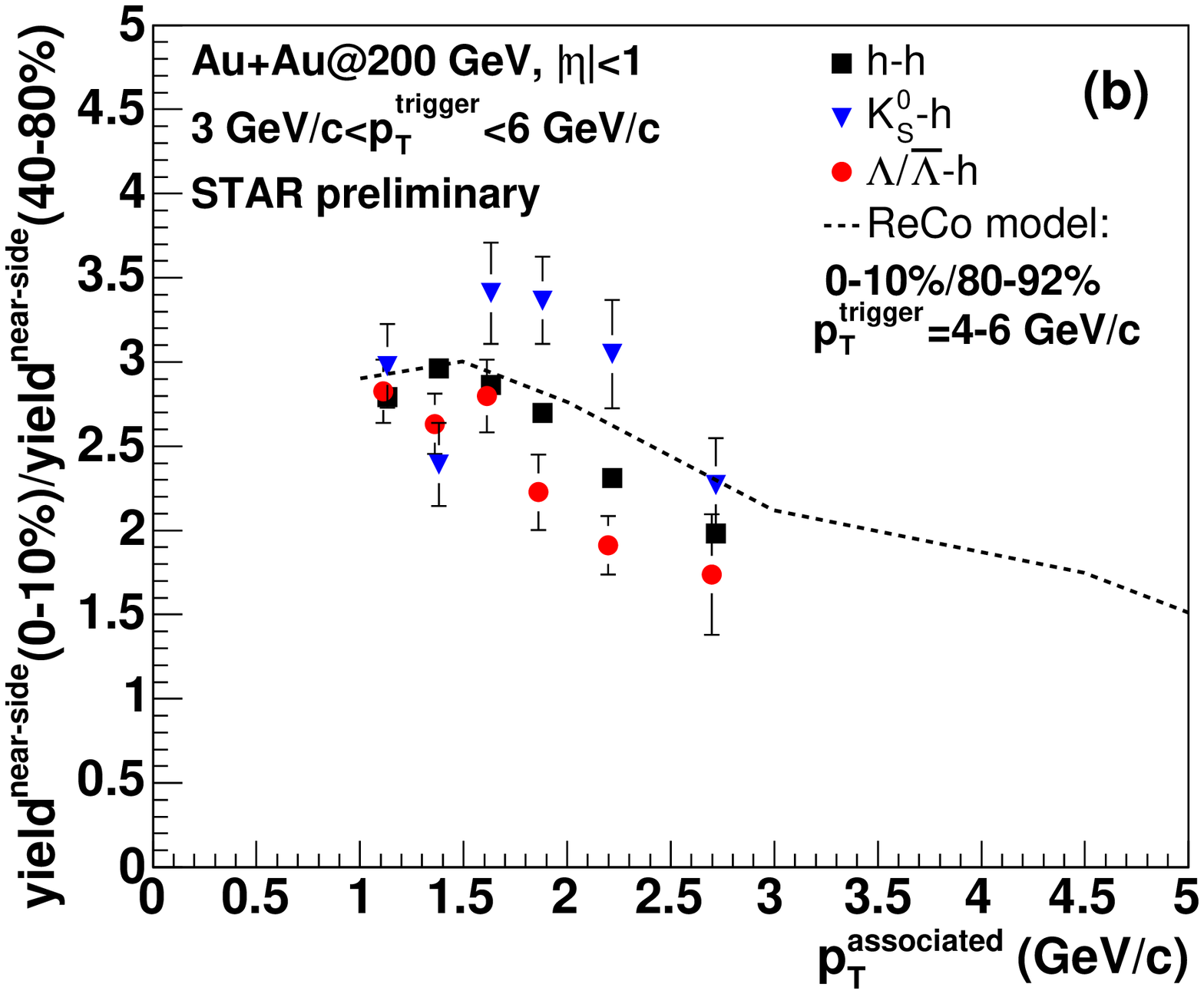}
\end{tabular}
\caption{(a) Invariant $p_T$ distribution of associated charged particles on the near
side in central Au+Au collisions. 
(b) The central-to-peripheral ratio of the near-side yield in  Au+Au collisions as a function of $p_T^{associated}$.
The dashed line is from~\cite{Hwa:2005ui}.}
\label{ptassochwa}
\end{figure}
Next, we study the dependence of the near-side yield on the transverse momentum of the trigger particle,
$p_T^{trigger}$, shown in Figure~\ref{rjpttrig}.
While the ridge yield  increases with $p_T^{trigger}$ and flattens off
for $p_T^{trigger}>$~3.0~GeV/$c$, the jet yield
keeps increasing with $p_T^{trigger}$. The jet yield for $\Lambda$ triggers
is systematically below that of charged and $K^0_S$ triggers.
Two effects could possibly explain this difference:
(1) the heavier baryon takes away more energy than the meson and thus
less energy is available for the associated particle production, (2)
an artificial track merging affects more $\Lambda$ than $K^0_S$ and charged particles
resulting in a lower associated yield.
Both effects are currently under investigation.

Fig.~\ref{ptassochwa}(a) shows the invariant $p_T$ distribution
of associated charged particles on the near side in central Au+Au collisions
for $p_T^{trigger}$~=~3-6~GeV/$c$. We have fit the data with
an exponential function $Ae^{-p_T/T}$ and extracted the inverse slope $T$.
The jet spectrum has  $T\sim$~520~MeV and is on average
80~MeV harder than the jet+ridge spectrum with $T\sim$~440~MeV.
In order to make a comparison with the parton recombination model~\cite{Hwa:2005ui},
we have calculated a central-to-peripheral ratio of the near-side yields in Au+Au as shown in
Figure~\ref{ptassochwa}(b).
The ratio is about 3 at $p_T^{associated}$~=~1~GeV/$c$ and decreases with $p_T^{associated}$.
Although the calculation has been done for charged pions, our results
qualitatively agree with the model and point toward a significant
role of thermal-shower recombination in  central Au+Au collisions.
To draw any quantitative conlusions, the calculation should be done for the same
centrality, $p_T^{trigger}$ selection and reproduce properties of the $\Delta\eta$ correlations.


In summary, we have reported recent results on strangeness production at intermediate-$p_T$
at RHIC. The strange baryon/meson ratios and identified particle $R_{CP}$ show that the
anomalous baryon production is limited to $p_T=$2-6~GeV/$c$ and above $p_T>$~6~GeV/$c$ both baryons
and mesons behave similarly. Two-particle correlations with identified strange trigger particles
reveal a strong contribution from the long-range pseudo-rapidity correlations at near-side.
Ongoing studies using identified associated particles will help to constrain the origin
of these correlations. The central-to-peripheral ratio of the near-side yields and
its decrease with $p_T^{associated}$ is in a qualitative agreement with the recombination model.
The crucial test of this model will be correlations with multiply-strange
trigger particles ($\Xi$, $\Omega$) which are under investigation.

\bibliographystyle{elsart-num}
\bibliography{biblio}
\end{document}